\documentclass[a4paper,11pt]{article}
\pdfoutput=1
\usepackage{comment}
\usepackage{jinstpub} 
\usepackage{color}
\usepackage{bxpapersize}
\usepackage{wrapfig}
\usepackage{multirow}
\usepackage{siunitx}

\title{Characterization of New Silicon Photomultipliers with Low Dark Noise at Low Temperature}

\author[a]{K. Ozaki,}
\author[b,c]{S. Kazama,}
\author[a,e]{M. Yamashita,}
\author[a,b]{Y. Itow,}
\author[d,e]{S. Moriyama}

\affiliation[a]{Institute for Space-Earth Environmental Research, Nagoya University, Nagoya, Aichi, 464-8601, Japan}
\affiliation[b]{Kobayashi-Maskawa Institute for the Origin of Particles and the Universe, Nagoya University, Nagoya, Aichi, 464-8601, Japan}
\affiliation[c]{Institute for Advanced Research, Nagoya University, Nagoya, Aichi 464-8601, Japan}
\affiliation[d]{Kamioka Observatory, Institute for Cosmic Ray Research, The University of Tokyo,Higashi-Mozumi, Kamioka Hida, Gifu 506-1205, Japan}
\affiliation[e]{Kavli Institute for the Physics and Mathematics of the Universe (WPI), The University of Tokyo,Kashiwa, Chiba 277-8583, Japan}

\emailAdd{nucr-pub1@isee.nagoya-u.ac.jp}

\abstract{Silicon photomultipliers (SiPMs) have low radioactivity, compact geometry, low operation voltage, and reasonable photo-detection efficiency for vacuum ultraviolet light (VUV). Therefore it has the potential to replace photomultiplier tubes (PMTs) for future dark matter experiments with liquid xenon (LXe). However, SiPMs have nearly two orders of magnitude higher dark count rate (DCR) compared to that of PMTs at the LXe temperature ($\sim$ 165 K). This type of high DCR mainly originates from the carriers that are generated by the band-to-band tunneling effect.
To suppress the tunneling effect, Hamamatsu Photonics K. K. have developed a new SiPM with lowered electric field strength. We have characterized its performance in a temperature range of 153 K to 298 K. We demonstrated that the newly developed SiPMs have 6--60 times lower DCR at low temperatures than that of the conventional SiPMs.}


\begin{document}
\maketitle
\flushbottom

\section{Introduction}
\label{sec:intro}
In the past three decades, numerous terrestrial experiments have been conducted to search for a faint interaction between weakly interacting massive particles (WIMPs) and ordinary matter. Among them, experiments using dual-phase (liquid/gas) xenon time projection chambers (TPCs) are leading the search for WIMPs with masses ranging from a few GeV/c$^2$ to a few TeV/c$^2$~\cite{XENON1Texposure,PandaX,LUX,XMASS}. For future experiments with larger detector mass such as DARWIN~\cite{DARWIN}, it is important to further decrease the backgrounds for reaching the sensitivity limited by atmospheric and supernova relic neutrinos (neutrino floor)~\cite{neutrino}. In the current experiments with  liquid xenon (LXe), photomultiplier tubes (PMTs) were used to detect the prompt primary scintillation (S1) with a wavelength of $\sim$ 175 nm~\cite{LXe} and secondary electro-luminescence of ionized electrons (S2), which are generated following an interaction between a WIMP and a xenon nucleus. However, PMTs have several important shortcomings, namely, their residual radioactivity levels~\cite{PMT-RI,XmassPMT}, bulkiness, and stability at cryogenic temperatures~\cite{XENON1T}. Therefore several alternative technologies are under consideration for future dark matter experiments with LXe.
Recently, Silicon photomultipliers (SiPMs), sensitive to vacuum ultraviolet light (VUV), have been developed by Hamamatsu Photonics K. K.~\cite{MEG2} and Fondazione Bruno Kessler~\cite{FBK,FBK_VUV}. It was reported that one of these SiPMs (Hamamatsu S13371~\cite{VUV4slide}) exhibited low intrinsic radioactivity~\cite{zurich}. Therefore SiPMs have the potential to replace PMTs for direct dark matter experiments. However, as listed in table~\ref{sipmpmt}, currently available SiPM has a high dark count rate (DCR) of 0.1--0.8 Hz/mm$^2$ at the LXe temperature ($\sim$ 165 K)~\cite{VUV}, which is $O$(10--100) times higher than that for PMTs (Hamamatsu R11410) used in the current dark matter experiments with LXe~\cite{PMTquarification}. 
Such a high DCR of SiPMs can generate numerous fake S1 signals due to accidental coincidences.
In the XENON1T experiment, 3-fold coincidence with 100~ns time-window is required to create S1 signals~\cite{daqsys}.
Assuming the same requirement is applied for a future multi-ton LXe experiment with 1900 PMTs~\cite{DARWIN}, this would result in an accidental coincidence rate of $O$(1) Hz for PMTs, which is similar to the fake S1 rate observed in the XENON1T experiment~\cite{acc}.
Therefore it is necessary to reduce the DCR of SiPMs at least to the level achieved for PMTs ($\sim$ 0.01 Hz/mm$^2$) for low energy threshold and background level~\cite{zurich2}.

\begin{table}[htbp]
\centering
\caption{\label{sipmpmt} Detector parameters for SiPMs and PMTs used in LXe experiments~\cite{VUV4slide, PMTquarification,VUV}.}
\smallskip
\begin{tabular}{|c|cc|}
\hline
Photo-sensor&SiPM&PMT\\
&S13370&R11410-21\\
\hline
Operation voltage&$\sim$ 50\ V&$\sim$ 1500 V\\
Single photoelectron gain&$\sim$ 2 $\times 10^6$ &$\sim$ 5 $ \times 10^6$ \\
Photo-detection efficiency at 175 nm&$\sim$ 24 $\%$ &$\sim$ 35 $\%$\\
DCR at LXe temperature ($\sim$ 165 K) &0.1--0.8\ Hz/mm$^2$&$\sim$ 0.01 Hz/mm$^2$\\
\hline
\end{tabular}
\end{table}

\section{New SiPM with lowered electric field strength}
\label{property}
We characterized a new SiPM (S12572-015C-SPL, which is hereafter referred as SPL) with a lower DCR developed by Hamamatsu as a prototype sensor. This SiPM is similar to the commercially available SiPM (S12572-015C-STD, which is hereafter referred as STD~\cite{STDSiPM}), but its internal electric field structure was modified to reduce the DCR as in ~\cite{FBK,Cryogenic_FBK}.
Dark pulses of SiPMs mainly originate from the carriers generated thermally~\cite{SHR} and by the band-to-band tunneling effect~\cite{tunnel}. The first component has a strong temperature dependence, while the second one has a weak dependence on temperature~\cite{cryo_study}. 
The high DCR of SiPM near the LXe temperature originates from the carriers due to the band-to-band tunneling, and the lowering of the internal electric field strength can suppress its effect, which enables to reduce the DCR \cite{SSPAD}.
Table~\ref{SPLSTD} shows the comparison of detector parameters for SPL and STD. 
From the table it can be seen that SPL and STD have the same properties of the active area, number of pixels, pixel pitches, and fill factor. 
Also, we can see from the table that operation voltage and single photoelectron (p.e.) gain are different because of the modified electric field structure~\cite{break}. It is noted that both SiPMs are not sensitive to VUV light.
The after-pulse probability of SPL is lower than that of STD because of lower doping concentration. Photo-detection efficiency of SPL is lower than that of STD due to the lower electric field strength depending on operation voltage. However, these differences are not significant for our purposes as described in appendix~\ref{APp} and \ref{relP}, therefore we will focus on the measurements of DCR and their results in the following sections.

\begin{table}[htbp]
\centering
\caption{\label{SPLSTD} Comparison of detector parameters between S12572-015C-SPL and S12572-015C-STD.}
\smallskip
\begin{tabular}{|c|cc|}
\hline SiPM&S12572-015C-SPL&S12572-015C-STD\\
\hline
Operation voltage&$\sim$ 100 V&$\sim$ 65 V\\
Gain at over-voltage = 6 V& 1.6 $\times$ 10$^5$ & 2.0 $\times$ 10$^5$\\
Active area&\multicolumn{2}{c|}{3 mm $\times $ 3 mm}\\
Number of pixel&\multicolumn{2}{c|}{40000}\\
Pixel pitches&\multicolumn{2}{c|}{15 $\mu$m}\\
Fill factor&\multicolumn{2}{c|}{53 \%}\\ 
Trench&\multicolumn{2}{c|}{No trench}\\ 
Spectral response range&\multicolumn{2}{c|}{320 -- 900 nm}\\ 
\hline
\end{tabular}
\end{table}

%%%%%%%%%%%%%%%%%%%%%%%%%%%%%%%%%%%%%%%%%%%%%%

\section{Experimental setup}
\label{exset_up}
\begin{figure}[]
    \centering
    \includegraphics[width=11cm]{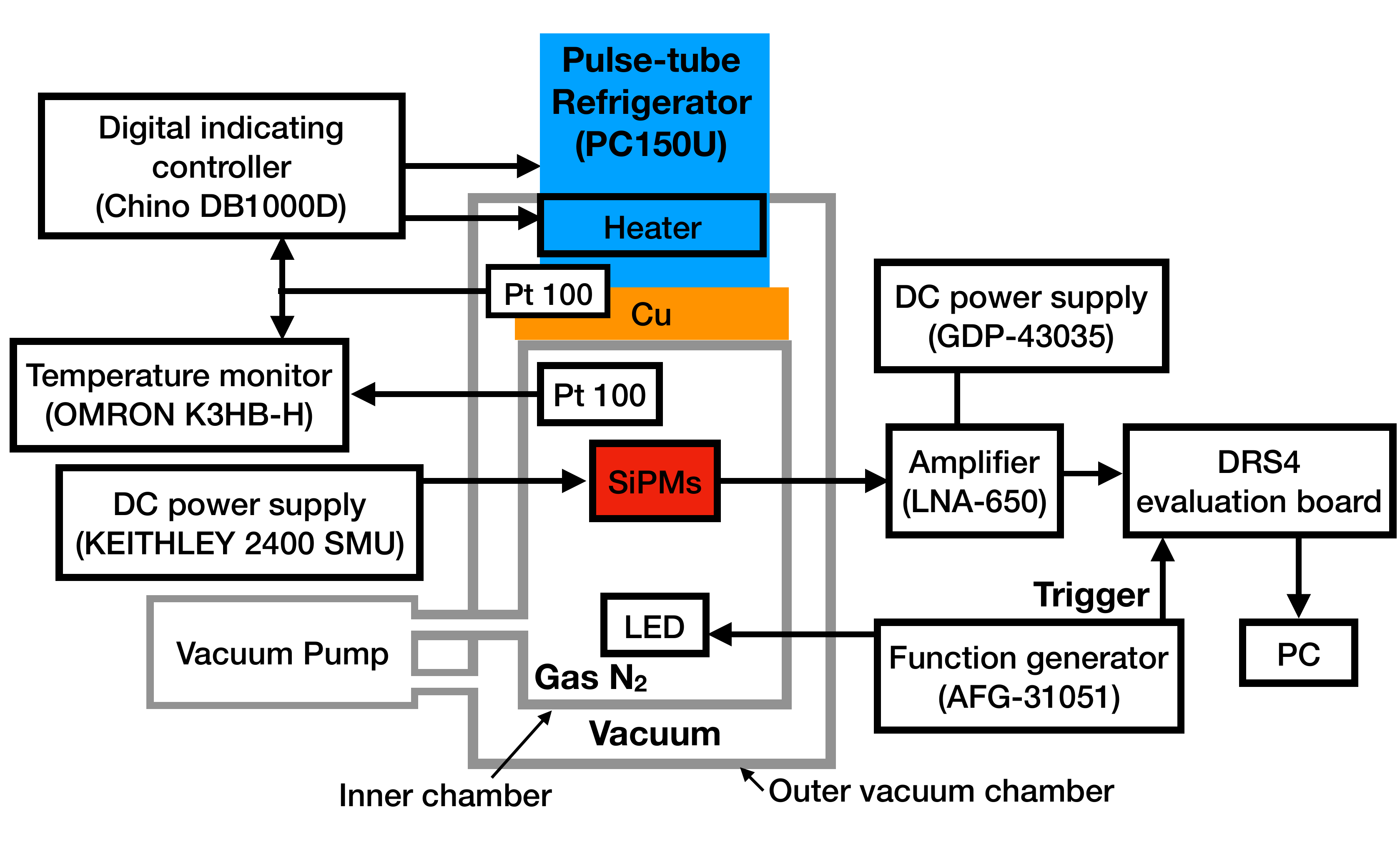}
    \caption{Block diagram of the setup of this experiment.}
    \label{block}
\end{figure}

\begin{figure}[]
\centering 
\includegraphics[width=7.0cm]{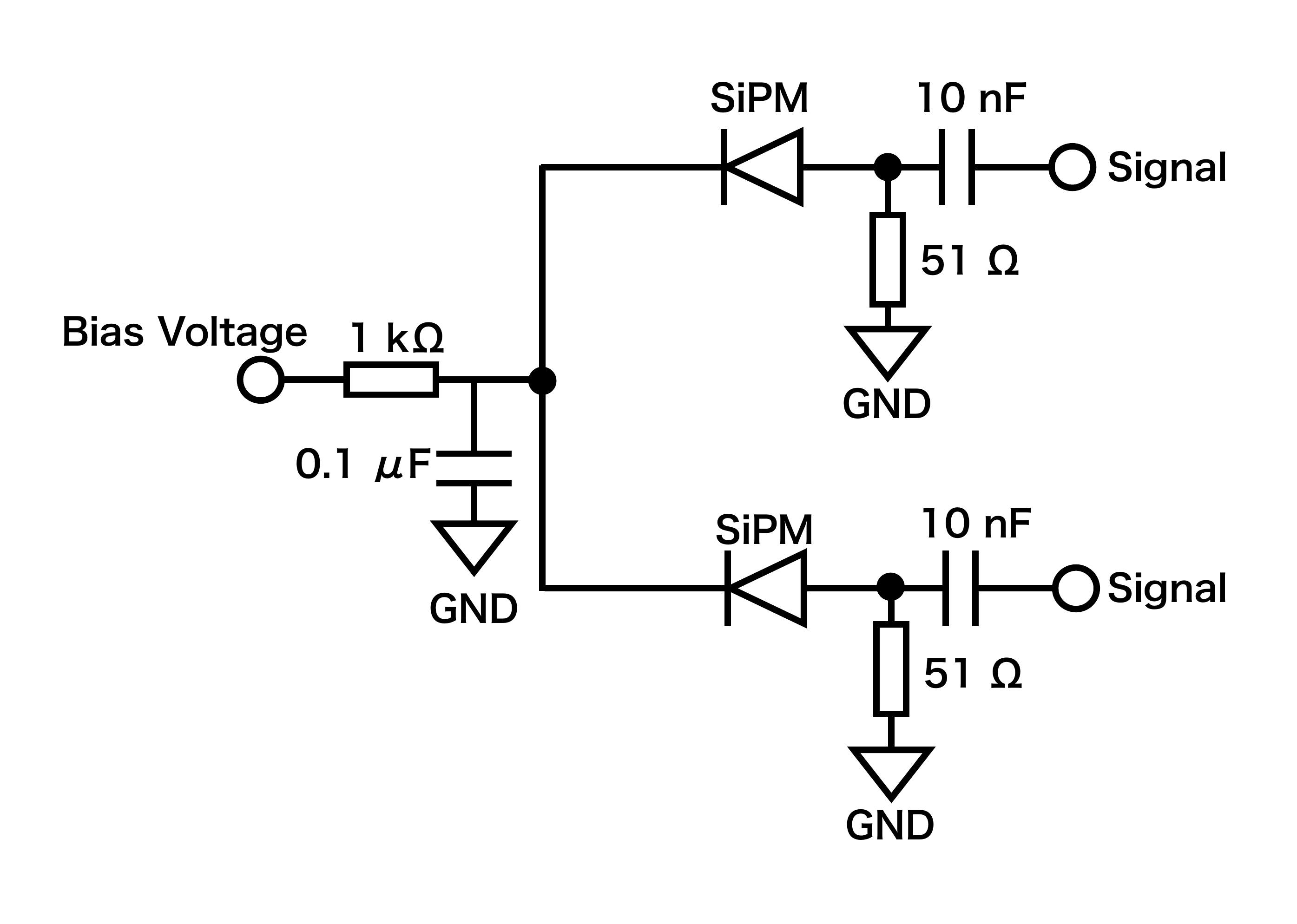}
\qquad
\includegraphics[width=7.0cm]{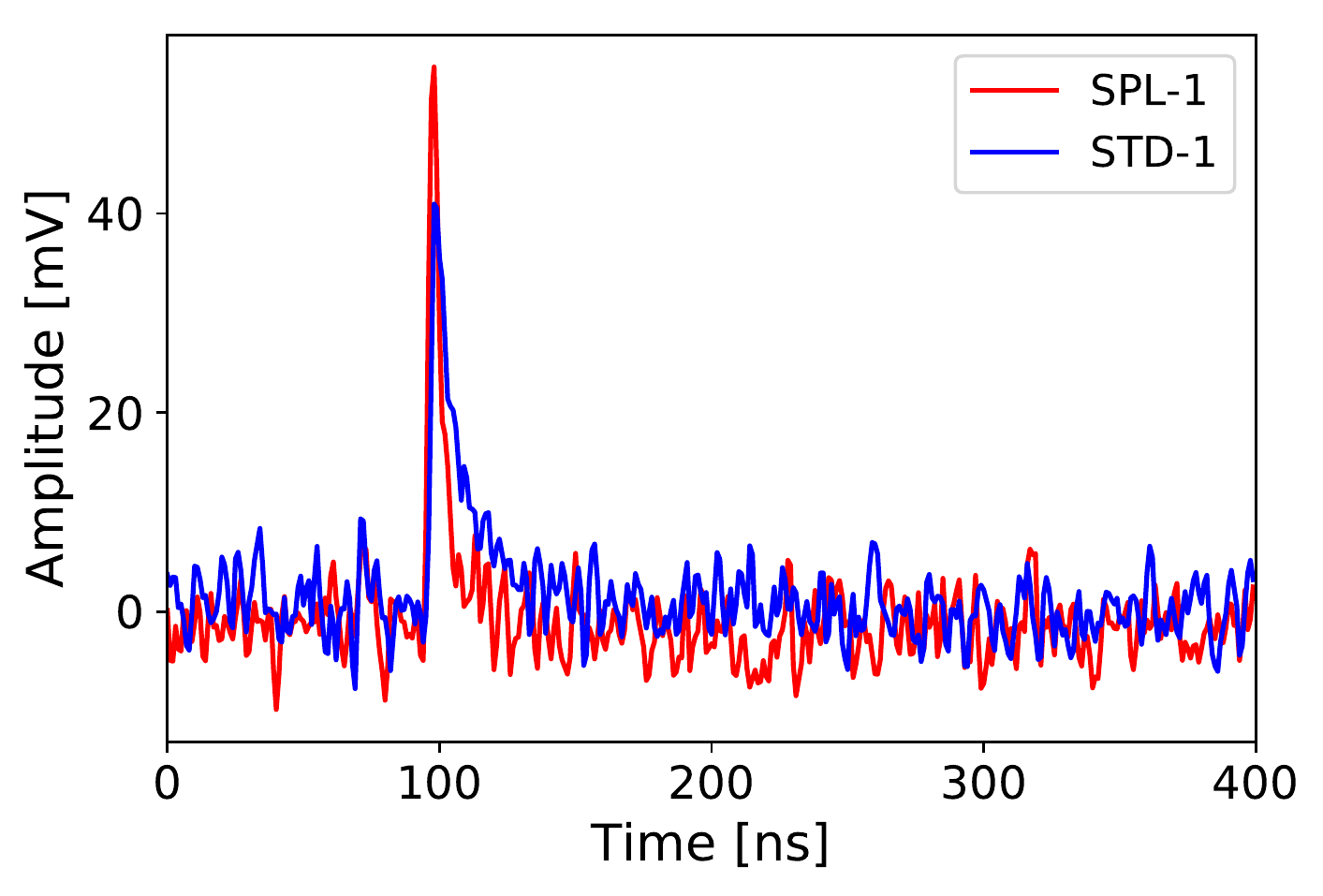}
\caption{\label{readout} (Left): Readout schematic of SiPMs. It is possible to readout two SiPMs using a common bias voltage. (Right): Typical waveforms of single p.e. for SPL (red) and STD (blue) at a temperature of 163~K and 164~K, and a bias voltage of 94.6 V (SPL) and 60.0 V (STD), respectively.}
\end{figure}

The experiment was performed in the Kamioka underground laboratory at a depth of approximately 1,000 meters. 
Figure~\ref{block} shows a block diagram of the setup that was used for this experiment. This setup consisted of an inner chamber filled with gas nitrogen along with an outer vacuum chamber. SiPMs were installed in the inner chamber and on a readout board, and the readout schematic is shown in figure~\ref{readout}, left. Voltage (V$_{\mathrm{bias}}$) from a source meter (KEITHLEY 2400 Source Meter) was applied to the SiPMs. Signals were amplified with a low noise amplifier (RF Bay LNA-650) with a 50db gain and read out using a DRS4 evaluation board~\cite{DRS4}, which acquired data with a rate of 1 GS/s and a length of 1024 data points per time window.
Square pulses from a function generator (Tektronix AFG-31051) were used to turn on an LED and trigger the data acquisition via the DRS4 board. We operated two SPLs labeled as SPL-1(-2) and two STDs labeled as STD-1(-2). Figure~\ref{readout} (right), shows the typical waveforms of SPL-1~(STD-1) acquired with DRS4 at a temperature of 163~(164)~K and bias voltages of 94.6~(60.0)~V. 
The temperature was measured by two Pt-100 sensors and was controlled with a pulse-tube refrigerator (ULVAC CRYOGENICS PC150U) and a heater connected to a digital indicating controller (Chino DB1000D). We operated the setup at a temperature range of 153 K to 298 K, which was stable within $\pm$ 0.5 K during the measurements.

\section{Measurements and Results}
\subsection{Gain vs voltages, breakdown voltage, and single photoelectron spectrum}
\label{sec:gain_measurements}
\begin{figure}[bhtp]
\centering
\includegraphics[width=7cm]{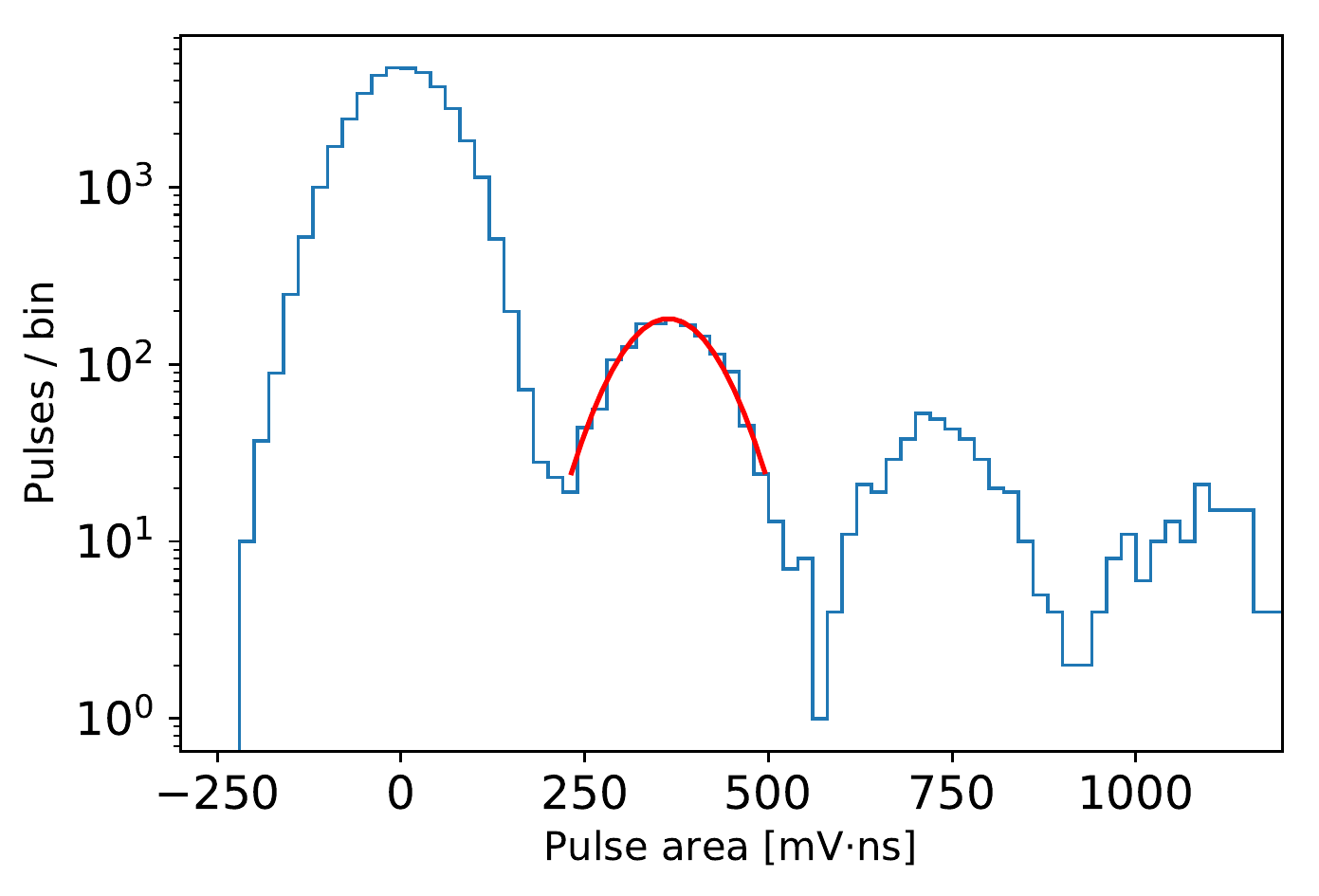}
\caption{Pulse area spectrum of SPL-1 at a temperature of 163~K and a bias voltage of 94.6~V with a fitted Gaussian function (red line). The first and second peaks correspond to pedestal and single p.e., respectively.}
\label{spe}
\end{figure}

To investigate the single p.e. gain, resolution, and breakdown voltage ($V_{\mathrm{br}}$) properties of both SiPMs, we measured their responses to the blue light~($\lambda\sim$ 375~nm) from an LED. The function generator inputs square pulses with a frequency of 460~Hz and width of 30~ns to the LED. We obtained 40,000 triggered events for each temperature and bias voltage.
Figure~\ref{spe} shows a pulse area spectrum of SPL-1 at a temperature of 163~K and a bias voltage of 94.6~V. 

\begin{figure}[]
\centering
\includegraphics[width=7.0cm]{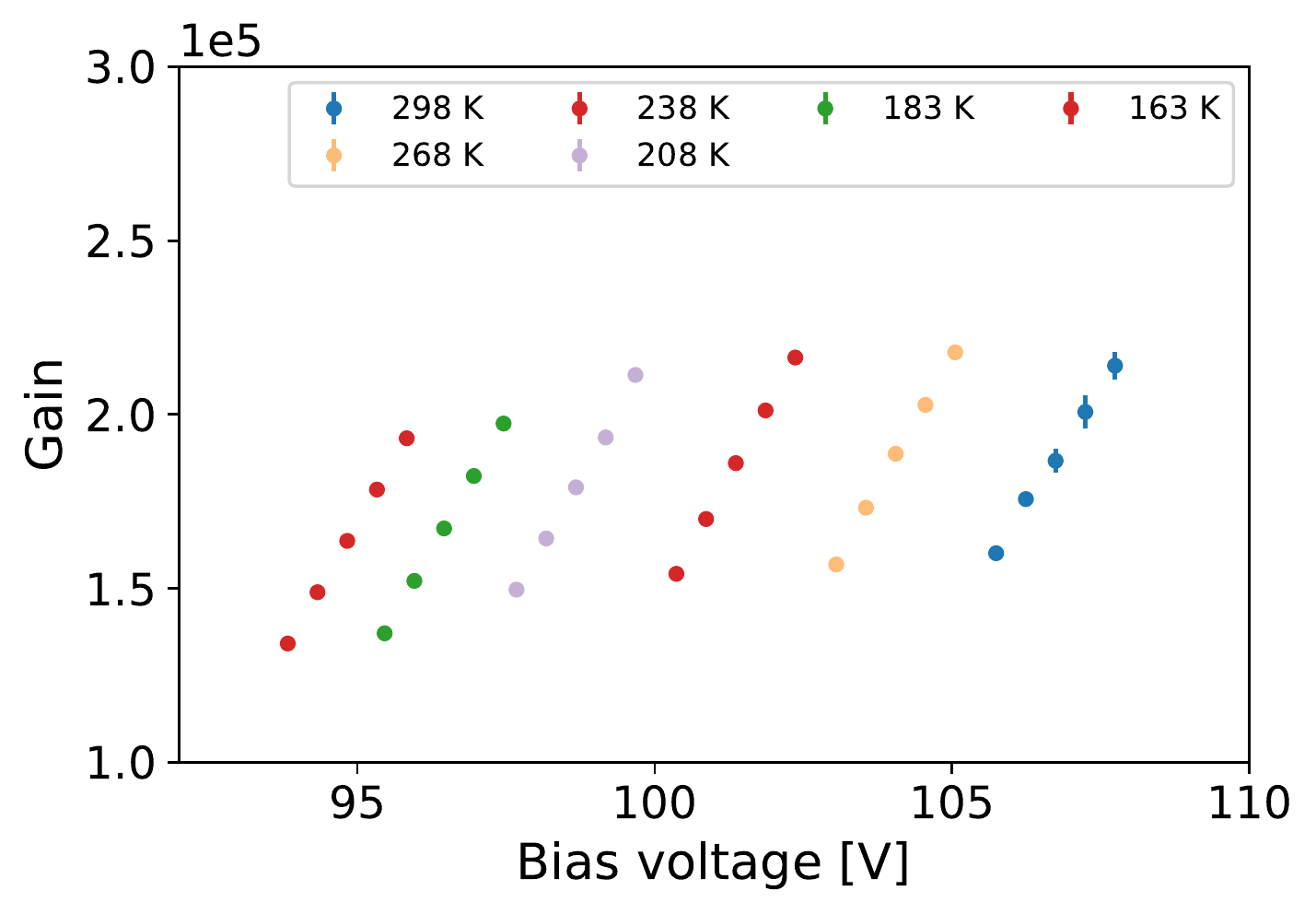}
\qquad
\includegraphics[width=7.0cm]{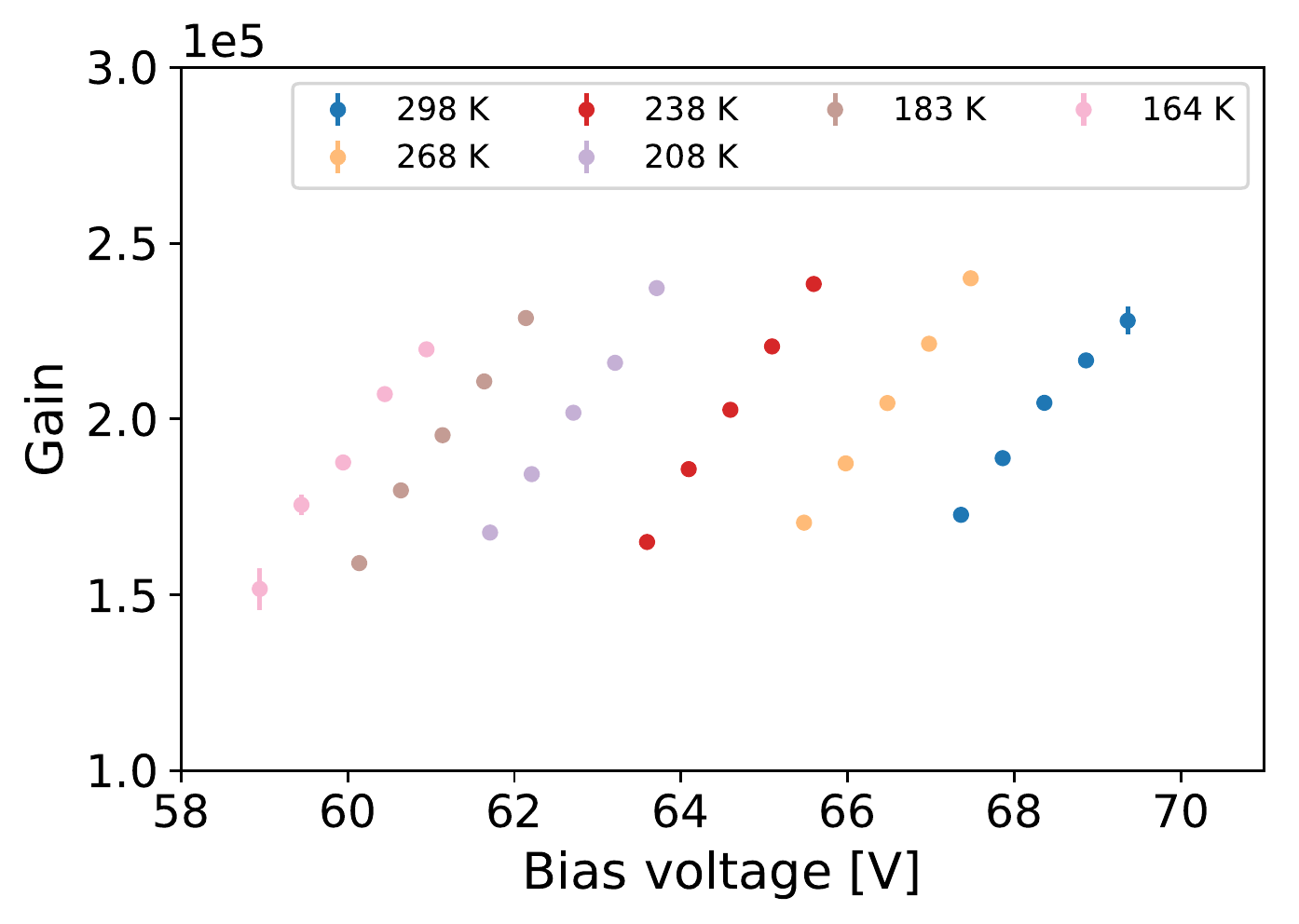}
\caption{\label{gain} Single p.e. gain of SPL-1 (left) and STD-1 (right) as a function of bias voltage for temperatures ranging from 153~K to 298~K.}
\end{figure}

\begin{figure}[bhtp]
\centering
\includegraphics[width=7cm]{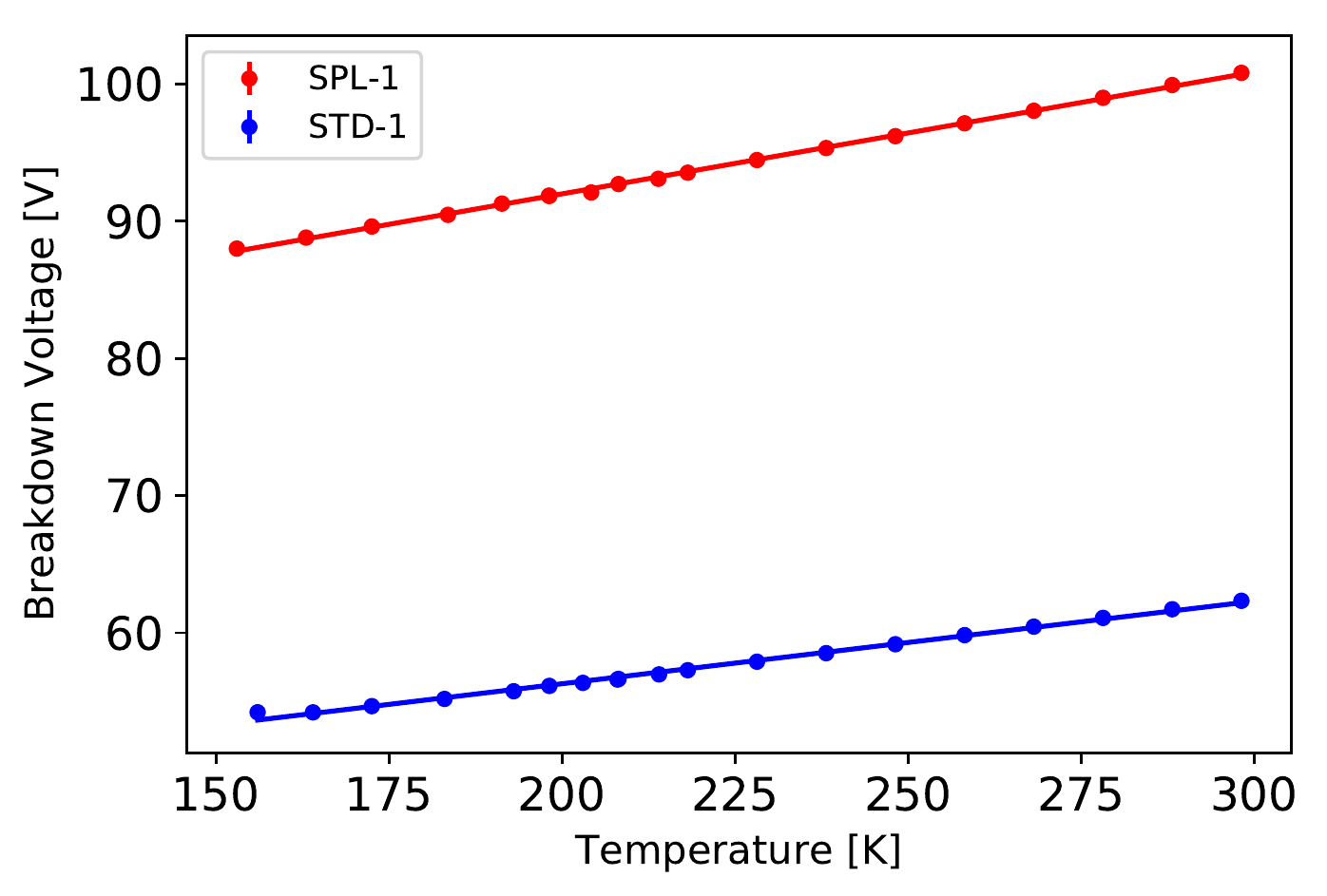}
\caption{Breakdown voltage of SPL-1 and STD-1 as a function of temperature.}
\label{break}
\end{figure}

At this temperature, single p.e. gains of SPL-1 and STD-1 at a bias voltage of 94.6~V (SPL) and 60.0~V (STD) are estimated to be 1.6~$\times$~10$^5$ and 2.0~$\times$~10$^5$, respectively.
The single p.e. resolutions ($\frac{\sigma_{1 p.e.}}{\mu_{1 p.e.}}$) of SPL-1 and STD-1 are measured to be 17.9~\% and 18.8~\%, respectively, where $\mu_{1p.e.}$ ($\sigma_{1p.e.}$) is mean (standard deviation) of the single p.e. pulse area obtained by fitting with Gaussian function as shown in figure~\ref{spe}. 
Figure~\ref{gain} shows a single p.e. gain of SPL-1 and STD-1 as a function of bias voltage. The single p.e. gain ($G$) can be expressed as
\begin{equation}
G=\frac{C_\mathrm{cell}(V_{\mathrm{bias}}-V_{\mathrm{br}})}{q},
\end{equation}
where $C_\mathrm{cell}$ is a cell capacitance and $q$ is an electric charge. The slope of a linearly fitted function in figure~\ref{gain} corresponds to $C_\mathrm{cell}/q$. The cell capacitance of SPL-1 and STD-1 are estimated to be 4.3~fF at 163~K and 5.4~fF at 164~K, respectively. 
Breakdown voltage ($V_{\mathrm{br}}$), where the gain collapses to zero, is shown in figure~\ref{break} as a function of temperature. We can see from the figure~\ref{break} that $V_{\mathrm{br}}$ of SPL-1 and STD-1 decrease with a slope of 88.3~mV/K and 61.2~mV/K, respectively. Also, approximately at the LXe temperature, $V_{\mathrm{br}}$ of SPL-1 and STD-1 is estimated to be 88.8~V and 54.2~V, respectively. 

Similar to low-field SiPMs developed by FBK~\cite{Cryogenic_FBK}, single p.e. gain of SPL is lower, and breakdown voltage and the slope in figure~\ref{break} of SPL is higher compared with STD. 
These observations can be explained by the fact that the depletion layer of SPL is thicker than that of STD as detailed in~\cite{Cryogenic_FBK}. 
For all the characteristics described above, no significant sample dependence is observed.

\subsection{Dark count rate and cross-talk probability}
\label{sec:dcr_measurements}

\begin{figure}[]
\centering
\includegraphics[width=7.0cm]{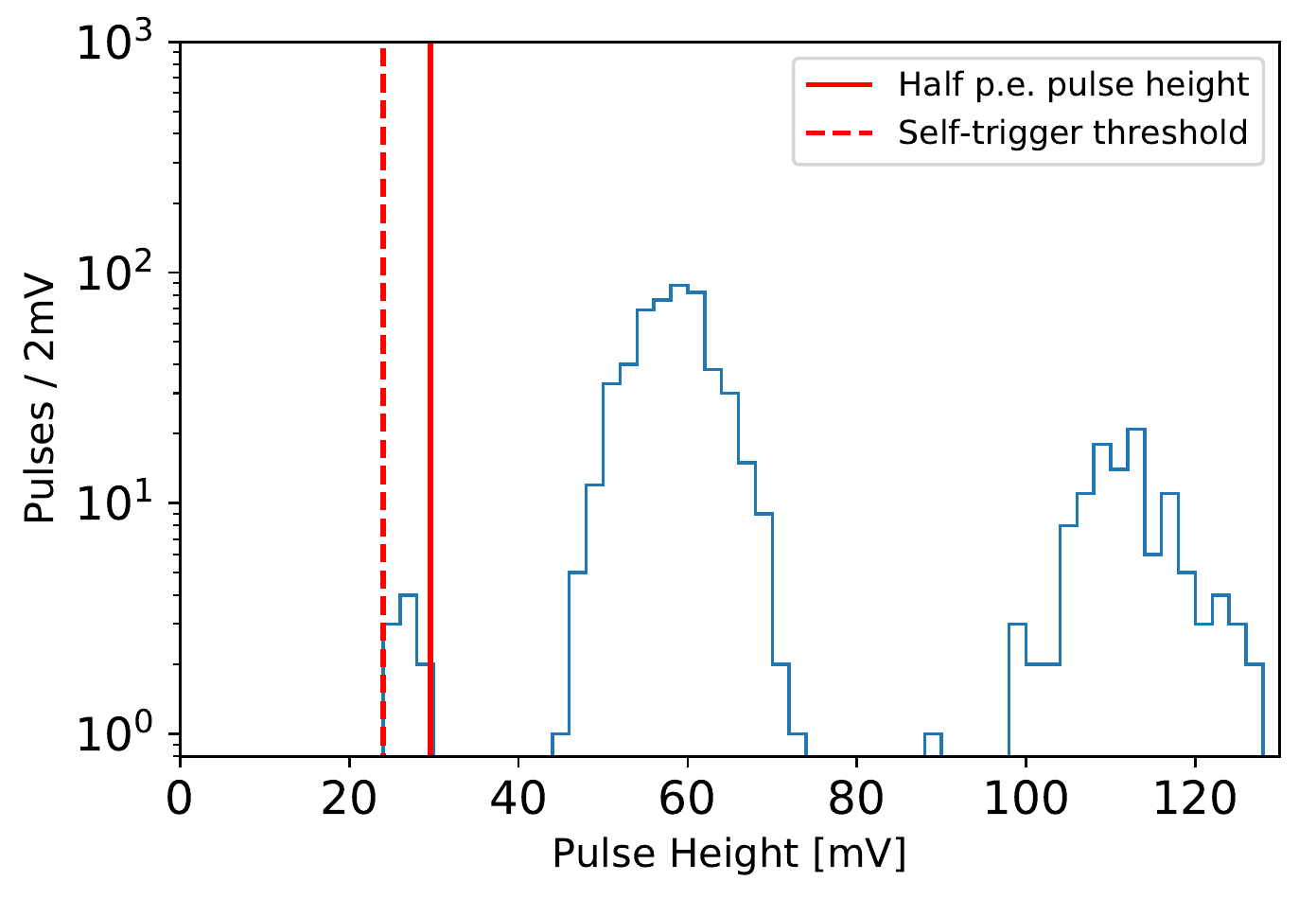}
\qquad
\includegraphics[width=7.0cm]{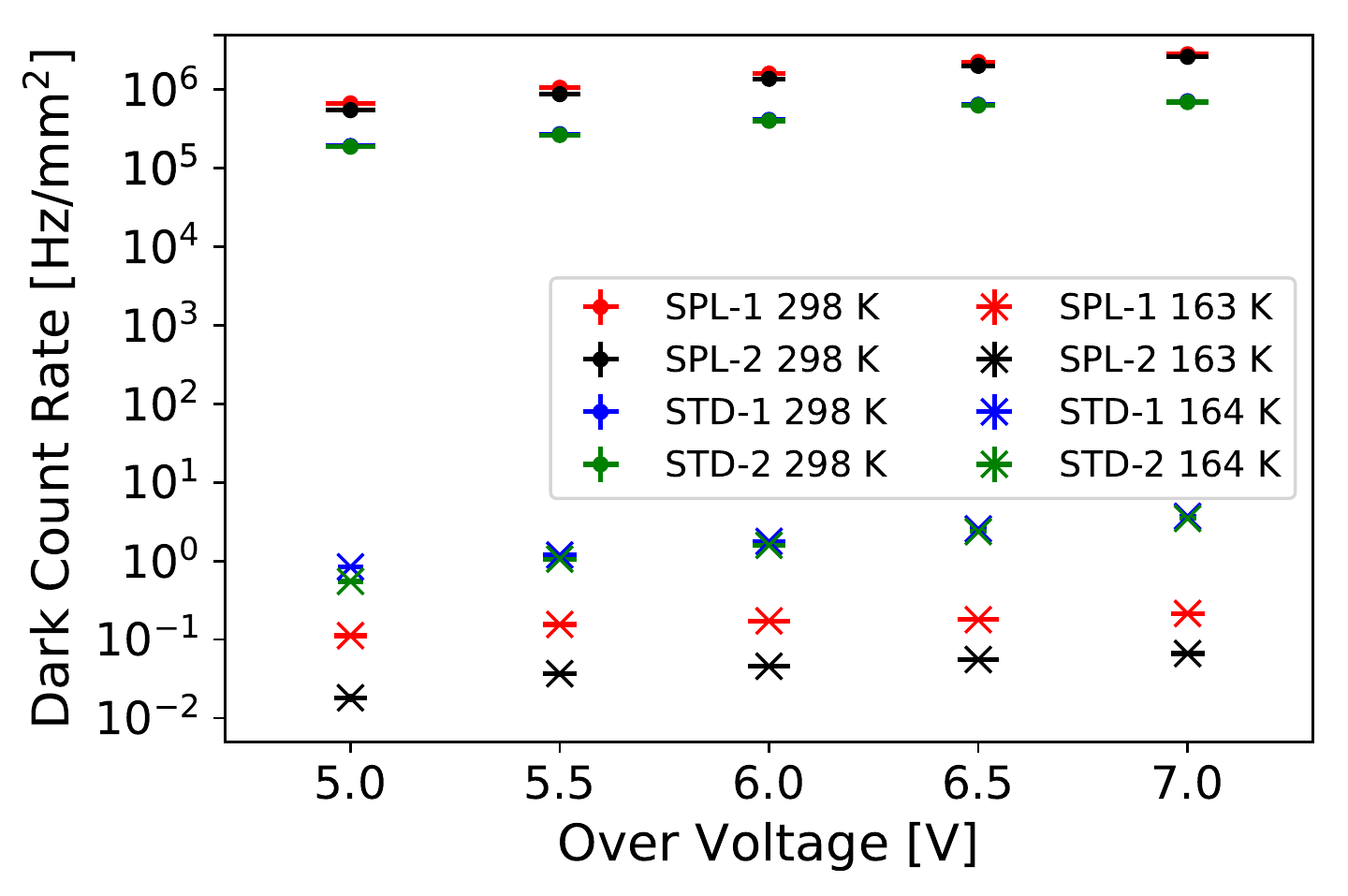}
\caption{\label{pulh} (Left): Pulse height distribution of SPL-1 acquired with self-trigger at a temperature of 163~K and an over-voltage of 6.0~V. Red dashed and solid lines show a self-trigger threshold of 25~mV and a half p.e. pulse height, respectively; (Right): Dark count rate of SPL and STD as a function of over-voltage at 298 K, 163 K~(SPL), and 164 K~(STD).}
\end{figure}

DCR is defined as the number of pulses per second whose height is larger than 0.5~p.e. In this study, it is normalized by the active area and the fill factor for each SiPMs. At a temperature between 198 K and 298 K, we acquired data with a random trigger because of high values of DCR. Below 198 K, data were obtained with a self-trigger with a threshold of 25~mV. Figure~\ref{pulh} (left) shows a pulse height distribution of SPL-1 at 163~K and an over-voltage of 6.0 V, where over-voltage is defined as $V_{\mathrm{bias}} - V_{\mathrm{br}}$. The peak around 60~mV corresponds to the mean of the single p.e. pulse height.
A red dashed line in figure~\ref{pulh} shows the self-trigger threshold of 25~mV. The remaining noise contribution is discarded by requiring that pulse area is larger than $\mu_{1p.e.}-3\sigma_{1p.e.}$. For the data acquired with self-trigger, deadtime of the DRS4 evaluation board ($\sim$ 3 \si{\milli s}) is taken into account when calculating the DCR.
Figure~\ref{pulh} (right) shows DCRs for SPL and STD that are expressed as a function of over-voltage at temperatures of 298 K, 163 K(SPL), and 164 K(STD).

\begin{figure}[]
\centering
\includegraphics[width=7.0cm]{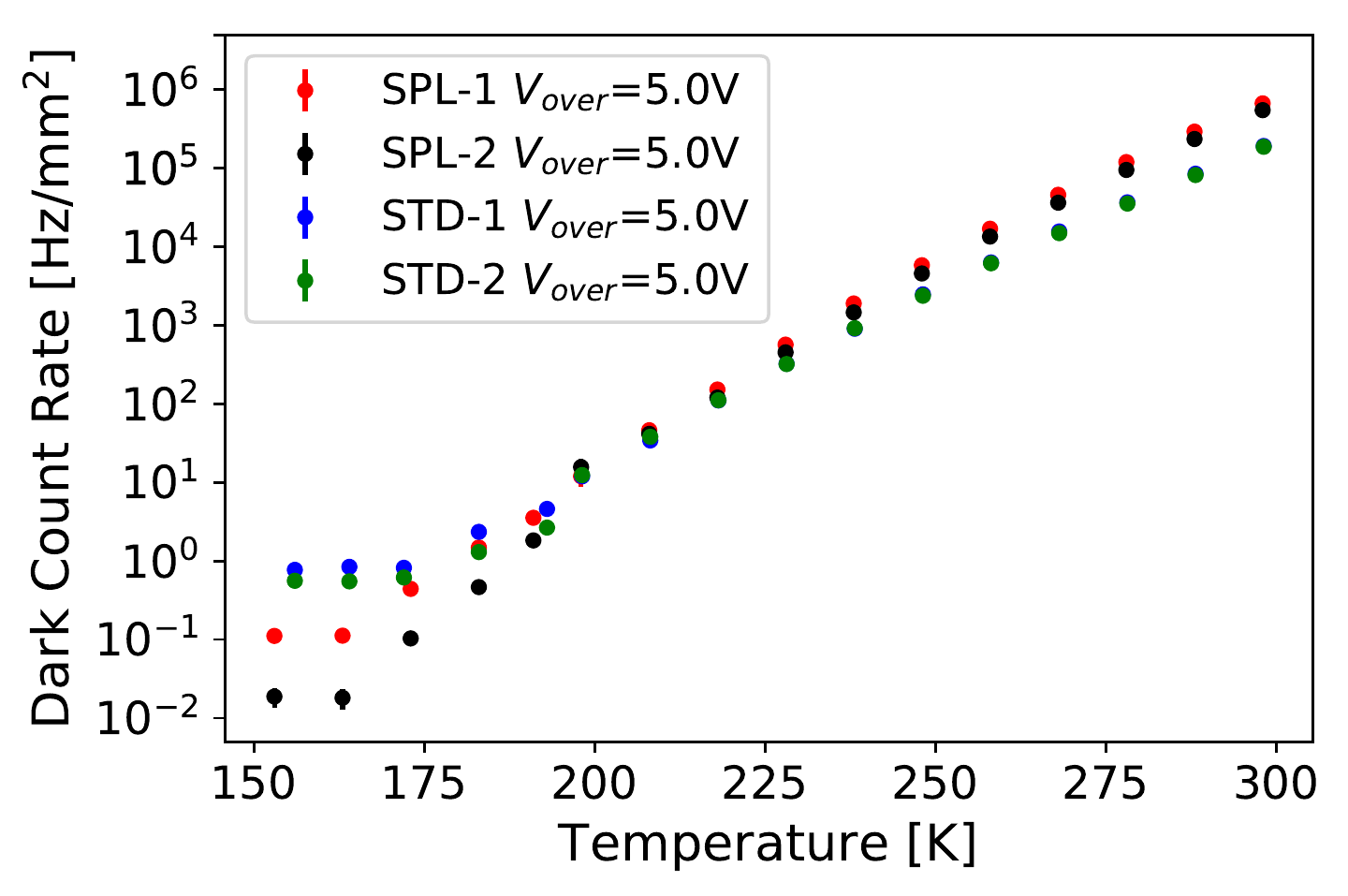}
\qquad
\includegraphics[width=7.0cm]{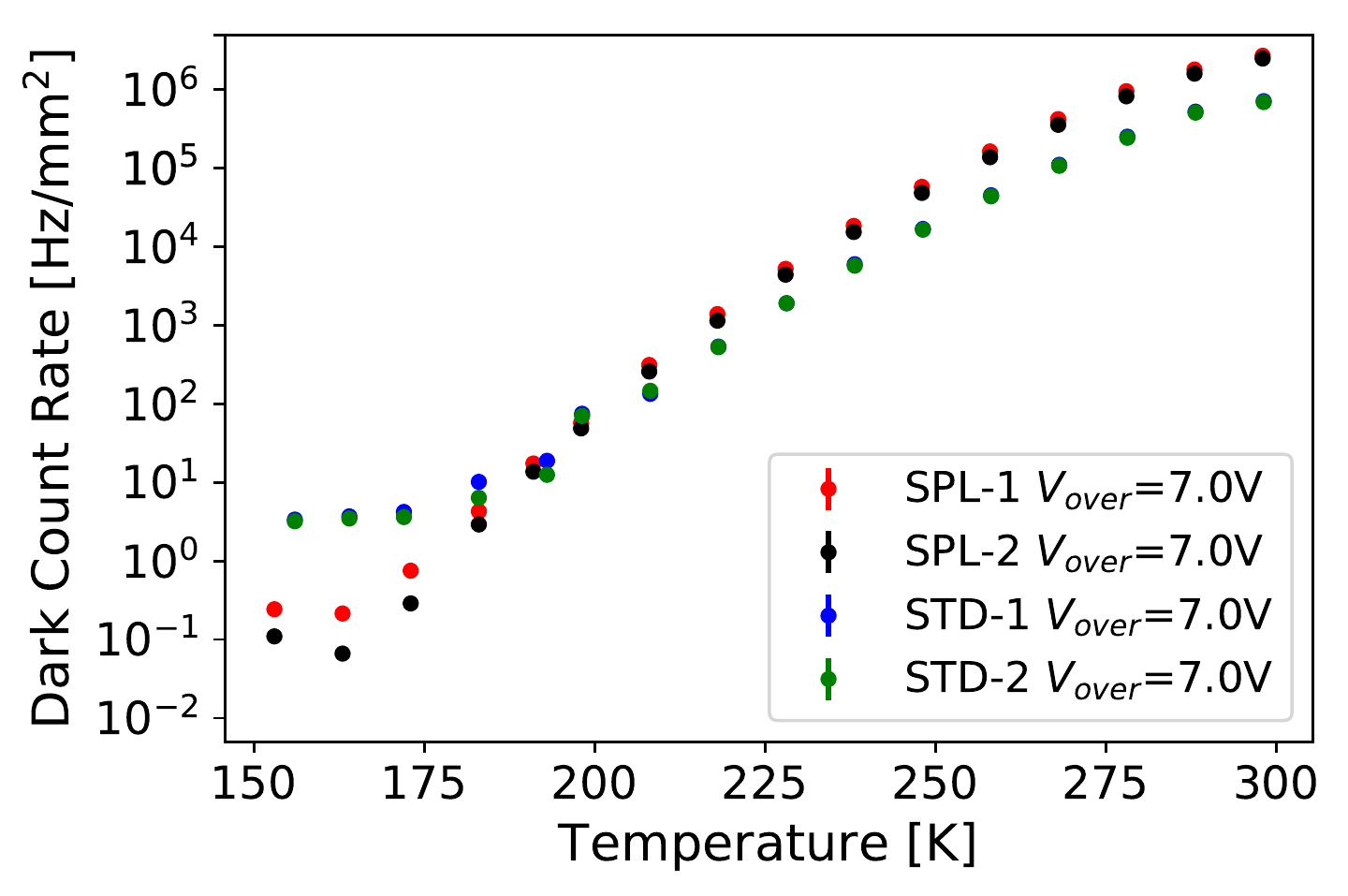}
\caption{\label{DCR} Dark count rate of SPL and STD as a function of temperature at over-voltages of 5~V (left) and 7~V (right).}
\end{figure}

DCR, as a function of temperature is also shown in figure~\ref{DCR} at over-voltages of 5.0 V and 7.0 V, respectively. As discussed in section~\ref{property}, between 200 K and 300 K, the contribution from thermally generated carriers is dominant; therefore, the DCR shows a rapid decrease with temperature. However, at temperatures below 200~K, the contribution from carriers originating from the band-to-band tunneling effect is dominant; therefore, the DCR has less temperature dependence. 
At room temperature, the DCR of SPL is found to be higher than that of STD. This can be originated from the fact that the depletion layer of SPL is thicker than that of SPL as discussed in section~\ref{sec:gain_measurements}. The DCR for SPL-1(-2) at approximately 165~K is measured to be 0.11--0.24~(0.018--0.071)~Hz/mm$^2$ depending on over-voltage, less by a factor of 6--60 compared with that of STD, indicating that the modified inner field structure reduced the DCR. 

\begin{figure}[]
\centering
\includegraphics[width=7.0cm]{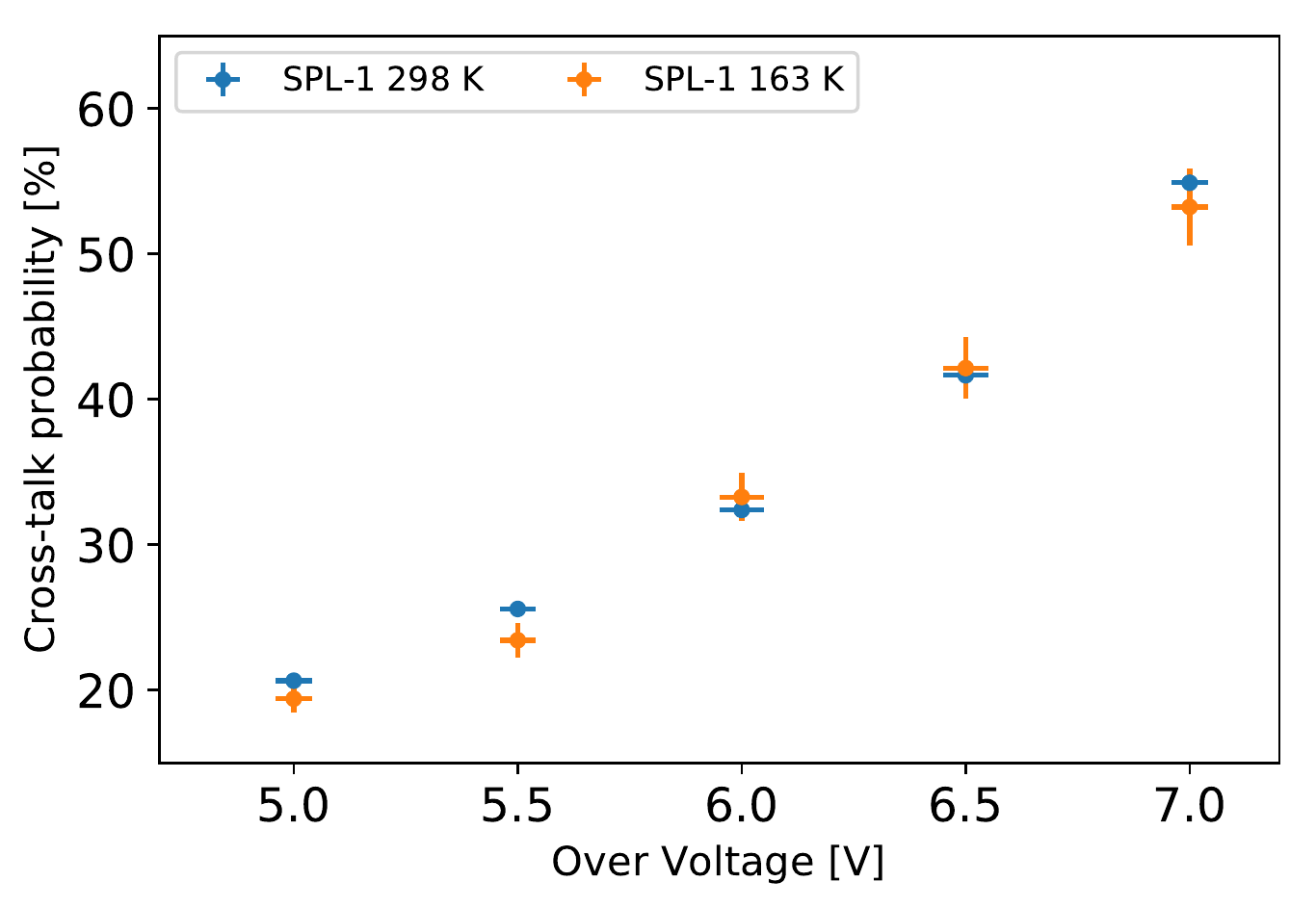}
\qquad
\includegraphics[width=7.0cm]{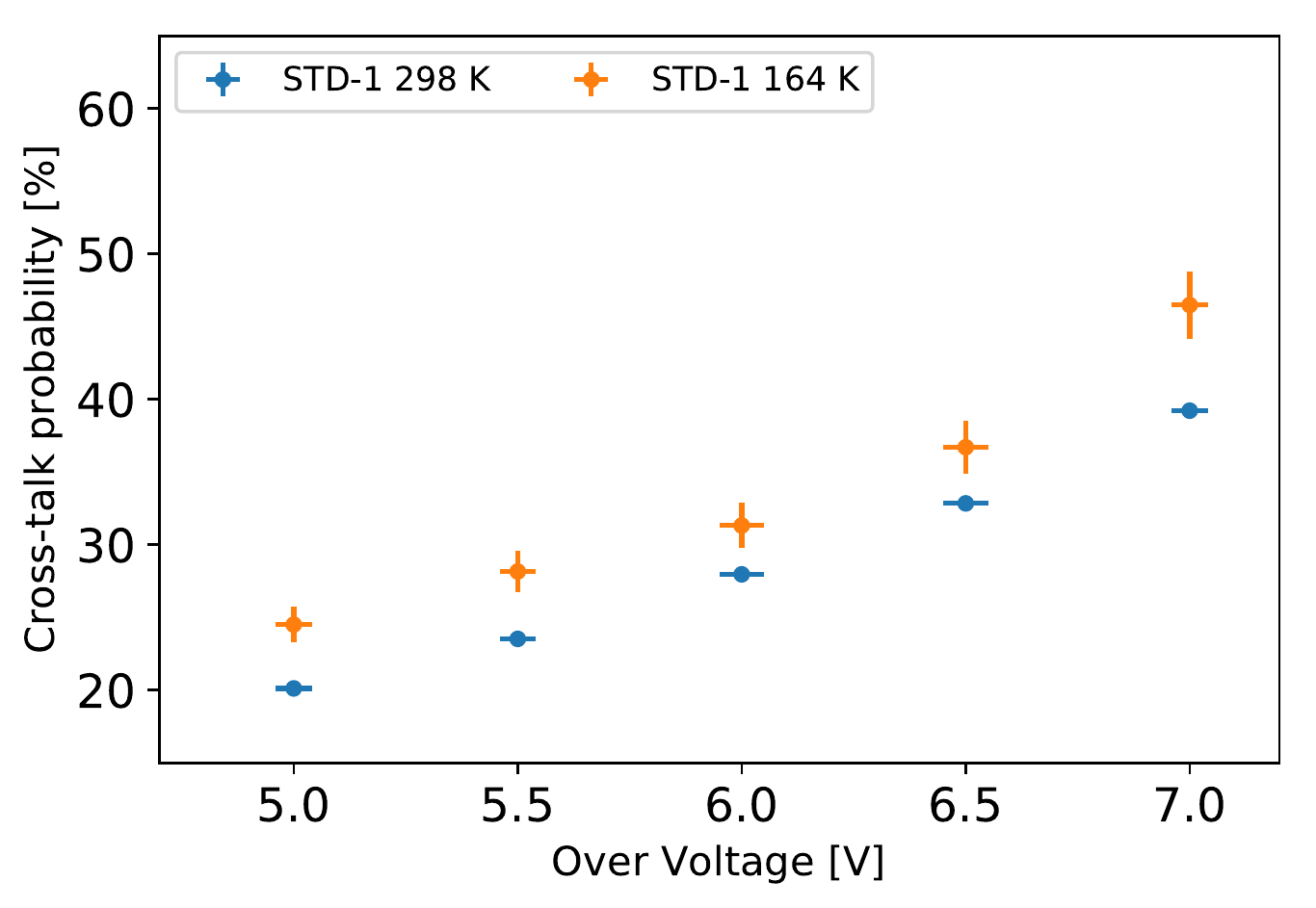}
\caption{\label{cross} Cross-talk probability of SPL-1 (left) and STD-1 (right) as a function of over-voltage.}
\end{figure}

Cross-talk probability (CTP) is calculated as $\frac{N_{1.5}}{N_{0.5}}$, where $N_{1.5}$ and $N_{0.5}$ are the number of pulses per second whose pulse heights are larger than 1.5~p.e. and 0.5~p.e., respectively~\cite{handbook}.
Cross-talk occurs when multiplied electrons in a cell enter and fire a neighboring cell.
Figure~\ref{cross} shows the CTPs for SPL and STD at different temperatures as a function of over-voltage. 
It can be seen from the figure~\ref{cross} that the CTP increases with over-voltage for both SiPMs. Moreover, the CTP of STD is found to increase with decreasing temperature. This is not well understood, but it can be because one of the cross-talk components, delayed cross-talk, has large temperature dependence as detailed in~\cite{decross}. 
At approximately the LXe temperature and an over-voltage of 6.0 V, the CTPs of SPL-1 and STD-1 are estimated to be 33.4~\% and 31.5~\%, respectively. No significant sample differences in the CTP are observed.

\section{Conclusion and Outlook}
SiPM is a good candidate for photo-detectors in future dark matter experiments with LXe because of its low radioactivity. However, it has a high DCR, and it is necessary to reduce it at least to the same level as it is for the PMTs ($\sim$ 0.01~Hz/mm$^2$) used in LXe experiments~\cite{PMTquarification}. Hamamatsu developed a dedicated SiPM (SPL) with the lowered electric field for suppressing the band-to-band tunneling effect, and we characterized it in a temperature range of 153--298~K. As a result, by modifying the inner electric field structure, the DCR at the LXe temperature could be reduced by a factor of 6--60 compared to that of STD that depended on over-voltage. 
Currently, both SPL and STD are not sensitive to the LXe scintillation light, but we are developing a dedicated SiPM sensitive to VUV light with the help of Hamamatsu. The DCR of S13370, a commercially available SiPM for VUV light detection, was measured to be 0.1--0.8~Hz/mm$^2$ at the LXe temperature depending on over-voltage~\cite{VUV}. With the same technology developed in this work, it might be possible that the DCR of S13370 can be reduced to the same extent, enabling SiPMs to be used in future dark matter experiments using LXe.

\acknowledgments
We thank Hamamatsu Photonics K. K. for this fruitful collaboration and the production of the SiPMs used in this study. We gratefully acknowledge the cooperation of Kamioka Mining and Smelting Company. This work was supported by DAIKO FOUNDATION, Foundation of Kinoshita Memorial Enterprise, the Japanese Ministry of Education, Culture, Sports, Science and Technology, Grant-in-Aid for Scientific Research, JSPS KAKENHI Grant Number 19H05805 and 20H01931, and the joint research program of the Institute for Cosmic Ray Research (ICRR), the University of Tokyo.

\appendix
%\section{Appendix}
\section{After-pulse probability}
\label{APp}
\begin{figure}[bhtp]
\centering
\includegraphics[width=7cm]{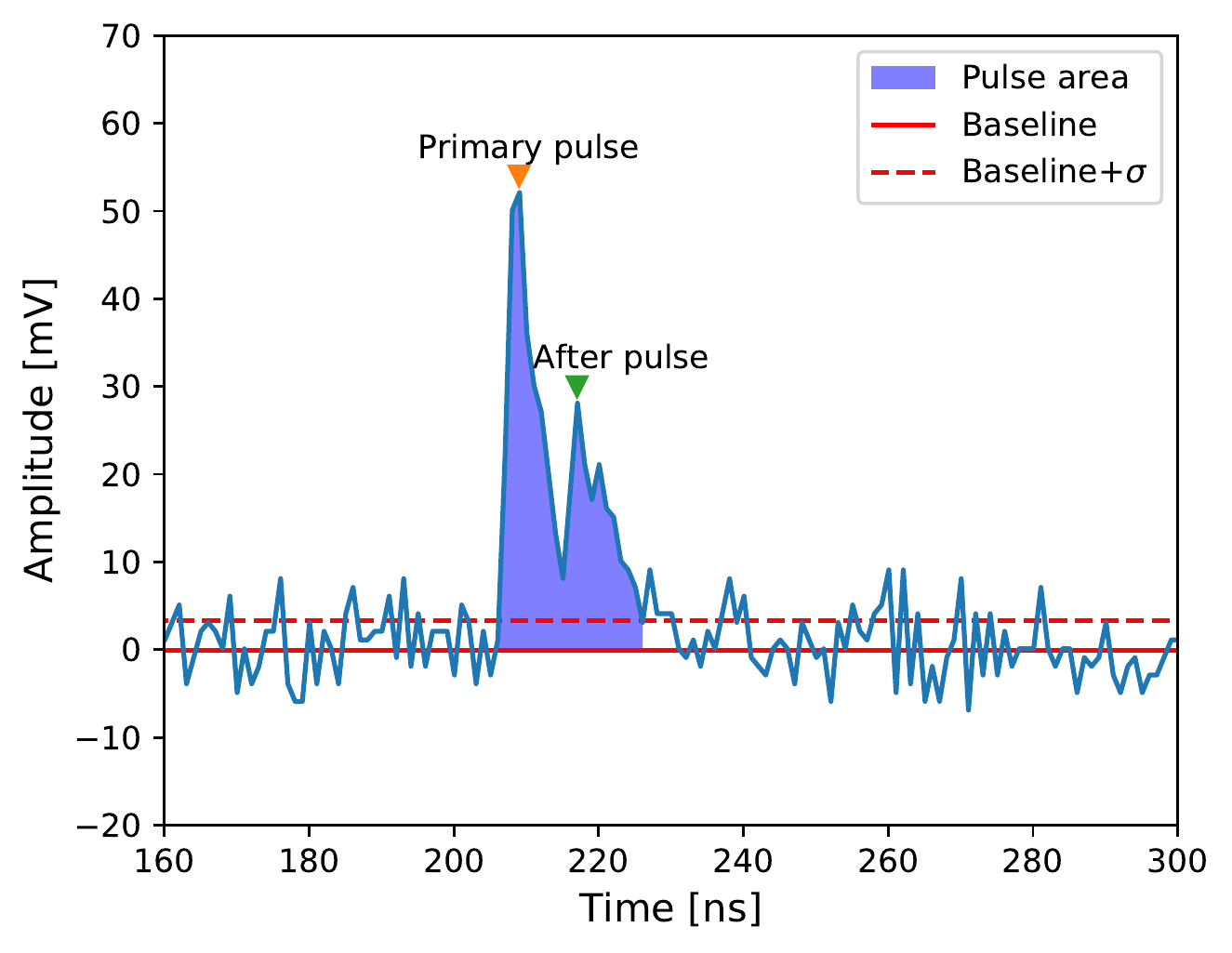}
\caption{Typical waveform of a single p.e. and an after-pulse. Blue shaded region corresponds to the region where pulse area is calculated.}
\label{APwaveform}
\end{figure}

\begin{figure}[]
\centering
\includegraphics[width=7.0cm]{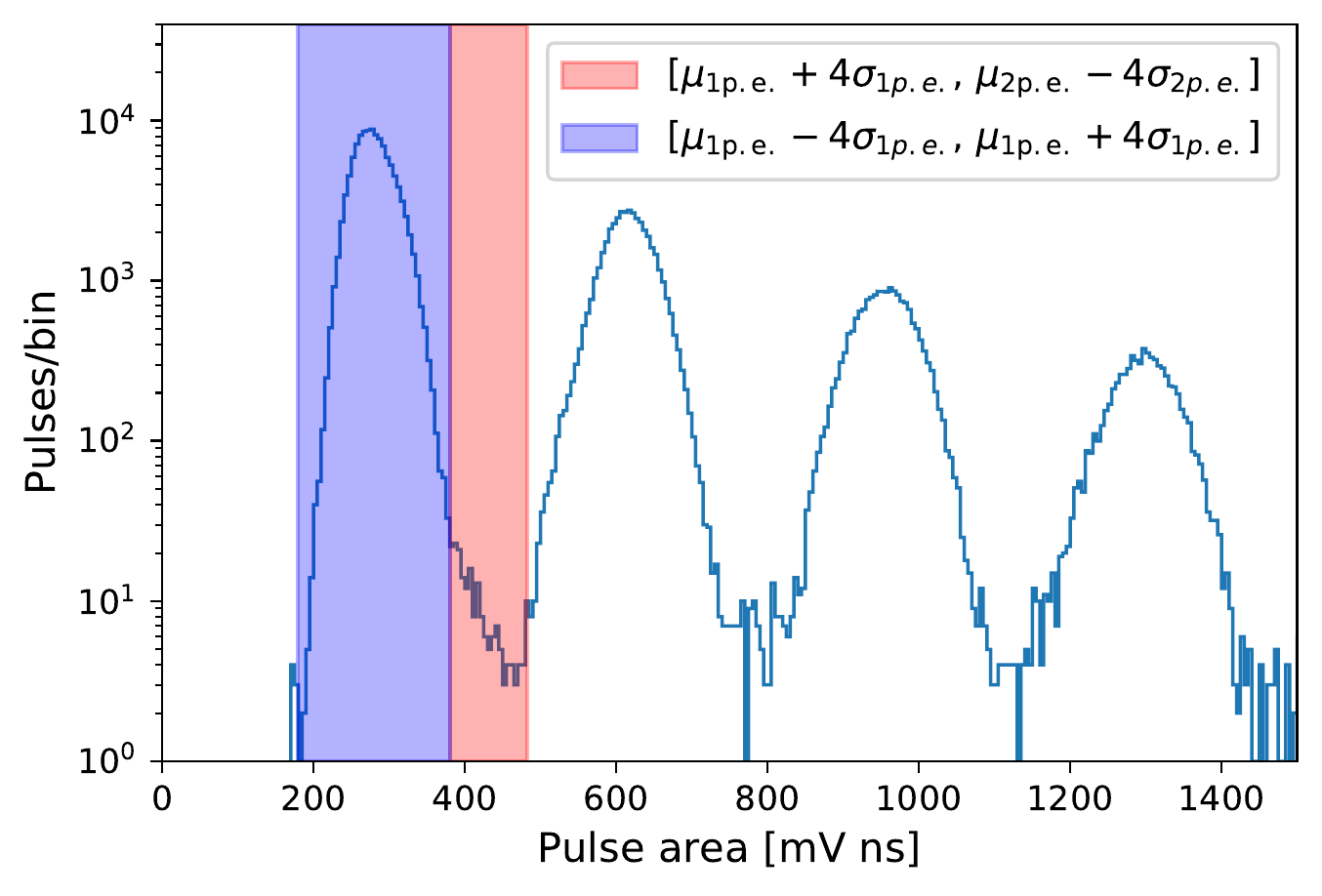}
\qquad
\includegraphics[width=7.0cm]{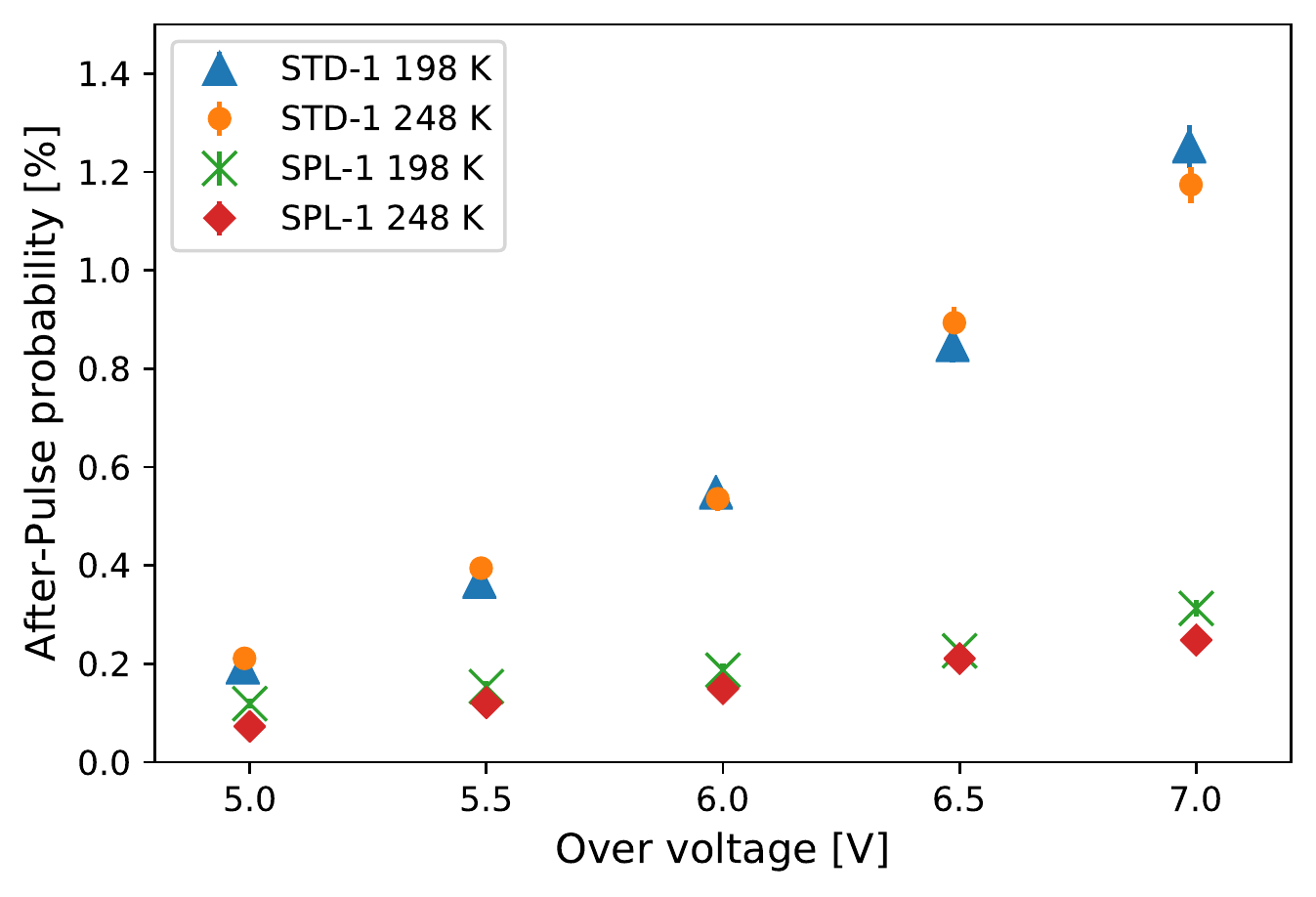}
\caption{(Left): Pulse area distribution of SPL-1 at a temperature of 198~K and bias-voltage of 97.0~V. (Right): After-pulse probability of SPL-1 and STD-1 as a function of over-voltage at temperature of 198~K and 248~K.}
\label{area_hist}
\end{figure}

The doping density of SPL is lower than that of STD because of the lowered electric field strength, therefore SPL is expected to have lower after-pulse probability compared with STD.
The number of carriers generated by the after-pulse process is less than that for single p.e., and an after-pulse signal generally piles up with a primary signal as shown in figure~\ref{APwaveform} since it is generated before the recovery time of the primary signal~\cite{approbability}.
Therefore, a single p.e. with an after-pulse signal has a pulse area ranging between single p.e. and double p.e.
Figure~\ref{area_hist} (left) shows the pulse area distribution of SPL-1 at a temperature of 198~K and bias-voltage of 97.0~V. The first and second peaks correspond to the pulse area for single p.e. and double p.e., respectively. Pulses with after-pulses are contained in the red shaded region defined by [$\mu_{\mathrm{1p.e.}}+4\sigma_{\mathrm{1p.e.}}$, $\mu_{\mathrm{2 p.e.}}-4\sigma_{\mathrm{2p.e.}}$].
In this study, after-pulse probability ($\mathrm{P}_{\mathrm{AP}}$) is defined as
\begin{equation}
\mathrm{P}_{\mathrm{AP}} = \frac{\mathrm{N}_{\mathrm{overlap}}}{\mathrm{N}_{\mathrm{1 p.e.}} + \mathrm{N}_{\mathrm{overlap}}},
\end{equation}
where $N_{\mathrm{overlap}}$ ($N_{\mathrm{1 p.e.}}$) is the number of pulses contained in the red (blue) shaded region.
Figure~\ref{area_hist} (right) shows the after-pulse probability of SPL-1 and STD-1 as a function of over-voltage.
As expected, the after-pulse probability of SPL is measured to be lower than that of STD, and no significant dependence on temperature has also been found.

%#\appendix
\section{Photo-detection efficiency}
\label{relP}
\begin{figure}[]
\centering
\includegraphics[width=6.0cm]{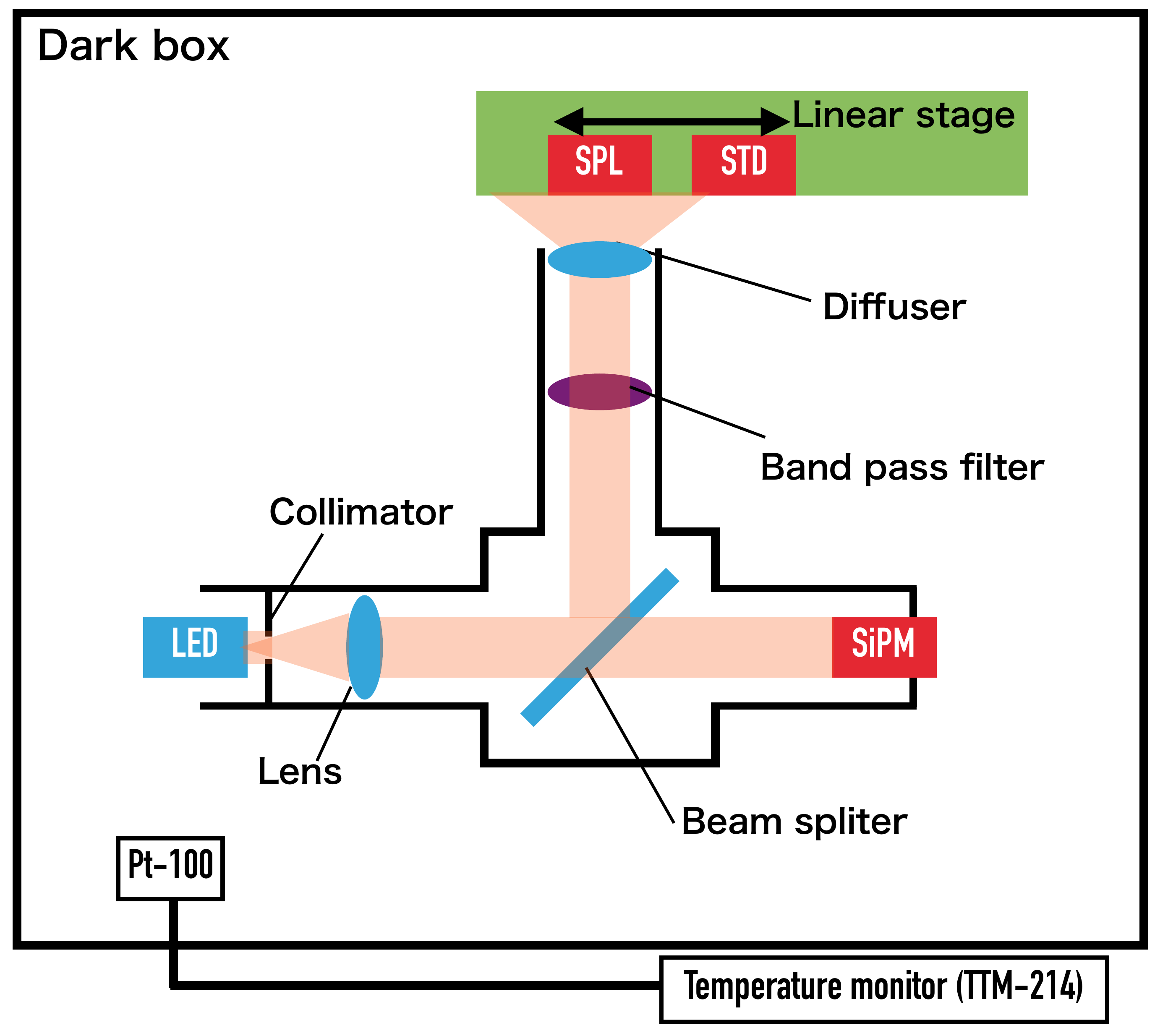}
\qquad
\includegraphics[width=8.0cm]{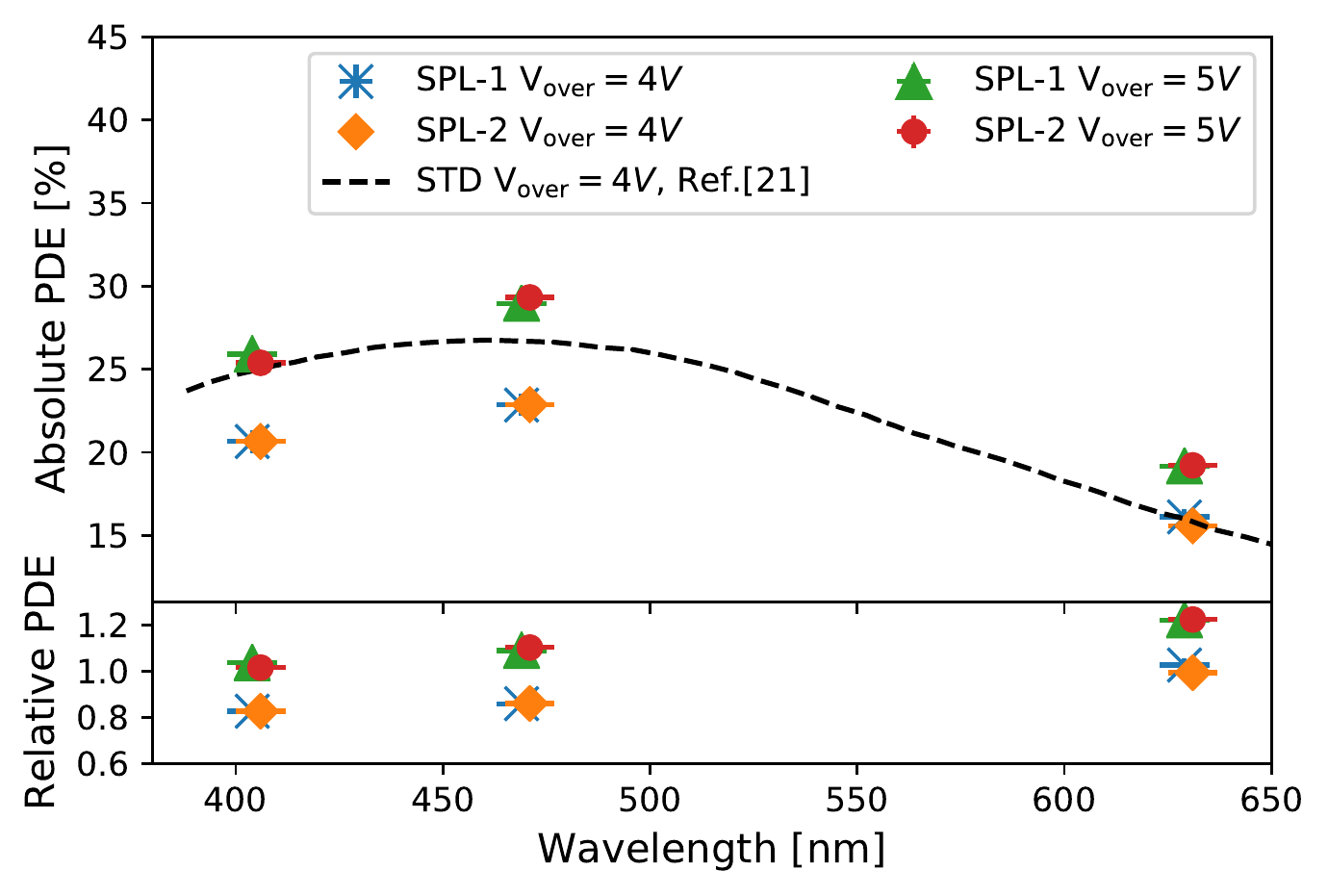}
\caption{(Left): Setup for the measurement of relative PDE. (Right): Measured PDE of SPL-1(-2) at over-voltages of 4~V and 5~V as a function of wavelength. Relative PDE to STD is also shown, where STD was operated at an over-voltage of 4~V. }
\label{rel_PDE}
\end{figure}

Since lowering the electric field strength can potentially reduce photo-detection efficiency (PDE), a comparison of PDE between STD and SPL has been done to investigate such an effect.
Figure~\ref{rel_PDE} (left) shows a setup to measure the relative PDE. 
Light emitted from an LED passes through a collimator with a diameter of 1 mm and a lens, and is then splitted with a beam splitter. The transmitted light illuminates the SiPM (Hamamatsu S13370-3050CS), and it is used to monitor the stability of the LED light. In this study, LED light level has been stable with 0.5~$\%$ fluctuation.
Reflected light passes through a band-pass filter and a diffuser, and then illuminates the SPL and STD. SPL and STD are mounted on x-y linear stage to control their positions remotely. Three LEDs with different wavelength of 405~nm, 470~nm, and 630~nm were used to investigate the wavelength dependence of PDE.  Temperature inside the bark box was monitored with a Pt-100 sensor, and had been stable at 297$\pm$1~K.
The relative PDE in this study is calculated by
\begin{equation}
\label{eq_pde}
\mathrm{Relative}\ \mathrm{PDE} = \frac{\lambda_{\mathrm{SPL-1(-2)}}}{(\lambda_{\mathrm{STD-1}} + \lambda_{\mathrm{STD-2}})/2.0},
\end{equation}
where $\lambda$ is occupancy for each SiPMs estimated with the method detailed in~\cite{withoutfit}. The absolute PDE is then calculated by scaling the estimated relative PDE by the absolute PDE of STD at an over-voltage of 4~V provided by Hamamatsu~\cite{STDSiPM}.
Figure~\ref{rel_PDE} (right) shows the measured (relative) PDE of SPL-1(-2) at over-voltages of 4~V and 5~V as a function of wavelength.
As can be seen from the figure~\ref{rel_PDE}, no significant differences in the estimated PDE between SPL and STD are observed. The relative PDE is found to have a wavelength dependence. This can be originated from the fact that both SiPMs have different depth profiles, and thus the sensitivity to VUV light can be affected by lowering the internal electric field strength, resulting in a reduction of PDE. However, a dedicated VUV-sensitive SiPM with higher fill factor is planed to be developed by Hamamatsu, which should improve PDE~\cite{hff}.

\color{black}

\end{document}